\documentclass[11pt]{article}
\usepackage{rotating}

\usepackage{lscape}

\usepackage{graphicx}
\begin{document}
\title{\bf Inconsistency of Chemical Properties of Stellar Populations in
the Thick Disk Subsystem of Our Galaxy}

\author{{V.\,A.~Marsakov,  V.\,V.~Koval', M.\,L.~Gozha,}\\
{Southern Federal University, Rostov-on-Don, Russia}\\
{e-mail:  marsakov@sfedu.ru, litlevera@rambler.ru, gozha\_marina@mail.ru}}
\date{accepted \ 2020, Astrophysical Bulletin, Vol. 75, No. 1, pp. 1-10}

\maketitle

\begin {abstract}

Using modern published data on velocities and spectroscopic 
definitions of chemical elements in stellar objects of the Galaxy, 
we investigated the relationship of chemical composition with the 
kinematics of different populations. The paper shows that the 
old stellar populations of the Galaxy, belonging (by the kinematic 
criterion) to the thick disk subsystem---globular clusters, field 
variables of the type RR Lyrae (lyrids), as well as close 
F--G dwarfs and field giants, have different chemical composition. 
In particular, the dwarfs and giants of the field are on average 
more metallic than the globular clusters and lyrids of the
field. Moreover, the relative abundances of $\alpha$-elements in 
the range $\rm{[Fe/H]} > -1.0$ are the highest for globular
clusters, and are the lowest for for field variables of the RR 
Lyrae type. Based on the analysis of the nature of the dependences 
of [$\alpha$/Fe] on [Fe/H] for these objects it was suggested 
that the thick disk subsystem in the Galaxy is composite and 
at least three components exist independently within it. The 
oldest one includes metal-rich globular clusters that formed 
from a single proto-galactic cloud shortly after the start 
of type Ia supernovae outbursts. Then the subsystem of field 
stars of a thick disk was formed as a result of ``heating''
of stars of already formed thin disk of the Galaxy by a 
rather massive dwarf satellite galaxy that fell on it. And finally, 
subsystems of field stars with the kinematics of not only 
a thick, but even a thin disk that fell on the Galaxy from 
this captured satellite galaxy.

\end{abstract}

{{\bf Key words:} Galaxy: structure--globular clusters: 
general--stars: variables: RR Lyrae}.

\maketitle

\section{Introduction}

As far back as the fifties of the last century, it was
noticed that metal-rich globular clusters occupy a
relatively small volume near the center of the Galaxy,
while metal-poor clusters are found in a much larger
space of the galactic halo (see, for example, Kinman
1959; Morgan 1959). The discussion continued
for several decades: are metal-rich clusters representatives
of the disk subsystem of the Galaxy, or is there
simply a negative radial gradient of metallicity in the
spherical galactic halo? It should be said that in
those years both the distances and the metallicities of
the clusters were determined with great uncertainties.
The existence of yet another subsystem intermediate
between the disk and spherical components was
first discussed after the appearance of the catalog
of globular star clusters (Kukarkin 1974) containing
the characteristics of 129 objects reduced to a single
system. An analysis of the catalog revealed a dip
of the metallicity function of clusters in the vicinity
of the value $\rm{[Fe/H]} \approx -1.0$, dividing all clusters into
two discrete groups (Marsakov and Suchkov 1976).
Moreover, the metal-poor group turned out to be a
spherically symmetric, slowly rotating halo subsystem,
and the metal-rich group was a rather flat fast-rotating
subsystem of the thick disk (Zinn 1985).
Since then, metal-rich clusters continue to be considered
as a separate subsystem, which was called the ``thick
disk''.

It turned out that the distribution of metals in
field stars is also discrete, and among them it is also
possible to distinguish an intermediate subsystem
by dips on the metallicity function (Marsakov and
Suchkov 1977). As a result, metallicity has become a
criterion for attribution of stellar objects to the Galaxy
subsystems, since it is a statistical indicator of age.
Indeed, in a closed star-gas system, which in the
first approximation can be considered our Galaxy, the
total abundance of heavy elements increases with time.
According to observations, the oldest stellar objects
in the Galaxy have the lowest metallicity, while stars
of solar age--the greatest. Therefore, despite contemporary
ideas about the formation of giant spiral
galaxies, like our own, from the merging of several
less massive ones in the early stages of evolution
of the Universe (see standard cosmological model
$\Lambda$CDM), the model of the monolithic collapse of a
protogalactical cloud described in the classical model
(Eggen~et\,al. 1962) has not lost its relevance. It is
clear that there is no single and sufficient criterion
for stratification. To reliably assign an object to a
particular subsystem, one should take into account
many parameters characteristic of each subsystem,
in particular, the position in the Galaxy, kinematics,
metallicity, the abundance of various chemical elements,
and age. There are no clear boundaries for the subsystems;
therefore, their sizes can be estimated only
approximately. Geometric boundaries imply certain
values and dispersions of velocities of objects belonging
to this subsystem. The use of kinematic
parameters is considered the most reliable method of
stratification of objects by subsystems. It is in this
way that close field stars are divided into subsystems
of the Galaxy. In particular, the technique described
in Bensby~et\,al. (2003) is widely used, when the probabilities
of belonging of nearby field stars to subsystems
of a thin and thick disks, or halo are calculated from
the components of residual velocities. It is understood
that the components of the spatial velocities of stars
in each subsystem obey normal distributions. The
average values of the velocity components, their dispersion,
and the relative proportions of stars in each
subsystem are set according to independent studies.
The subsystems differ mainly in the velocity of motion
around the galactic center: for a thick disk, it turns
out to be intermediate between the corresponding
velocities of a thin disk (approximately solar) and a halo
(almost zero). For objects distant from the Sun, the
velocity components must be in a cylindrical coordinate
system, since in the rectangular coordinate system,
the velocity components will be multidirectional
with respect to the center and direction of rotation of
the Galaxy, the values of which mainly determine their
subsystems.

The mechanism of formation of the thick disk subsystem
has long been a subject of discussion. The
modeling of the structures of our Galaxy is based on
the results of counting the field stars of both disk
subsystems and studying their chemical composition.
All the proposed scenarios for the formation of a
thick disk can be divided into several categories, but
all are in some way unsatisfactory. So, if a thick
disk is formed during the collapse of a protogalactic
cloud, then it will inevitably have a change in the
relative abundance of $\alpha$-elements with an increase in the
abundance of heavy elements (Prochaska~et\,al. 2000).
The difficulty of this representation is that the time
of complete collapse of the proto-galactic cloud is
much less than the characteristic time of evolution
of pre-supernova SNe Ia. This discrepancy is easily
avoided in another class of models, where the thick
disk subsystem is formed as a result of the heatihg 
of the primary stellar thin disk due to the
interaction of the Galaxy with a very close satellite
galaxy (see Kroupa~2002). The problem with this
model is to explain the existence of metal-rich globular
clusters in the thick disk. They cannot heat as
stars by the satellite galaxy and can arise only as a
result of star formation accompanying this interaction
(Gratton~et\,al. 2000). Another mechanism for the
formation of the subsystem involves the capture of
small dwarf satellite galaxies and getting their stars
in the thick disk (Abadi~et\,al. 2003). The formation of
a stellar thick disk inside our Galaxy is possible as a
result of the so-called ''wet merging'' with a satellite,
with a large amount of interstellar matter and the
accretion of its gas (Brook et al. 2004). Finally, a
thick disk could be produced from field stars simply
because of the radial migration of thin disk stars
(see Sch\"onrich and Binney 2009). Unfortunately, as
noted above, people try to explain the origin of the
thick disk, taking into account the detailed chemical
composition, only for stationary stars of the galactic
field, whereas for globular clusters and field variable stars
of the type RR Lyrae, such models do not take
into account the features of the abundances of chemical
elements in them.

This paper is a continuation of the study of chemical
and spatial-kinematic properties of globular clusters
and the field variables of the RR Lyrae type,
started in Marsakov~et\,al. (2018, 2019b,c). In them,
we, in particular, showed that the kinematics of both
types of objects do not combine well with metallicity,
therefore, to separate thick-disk objects from halo, we
should be guided for clusters by a dip, and for field
lyrids--by a knee of their metallicity functions in a
neighborhood of $\rm{[Fe/H]} \sim -1.0$
\footnote{Actually, the dip in globular clusters is located at a slightly
higher metallicity (see Borkova and Marsakov (2000)), but
it’s more convenient to accept this round number used commonly.}. 
Here we will compare
the chemical composition of globular clusters
and two types of field stars (close F and G dwarfs and
giants and stars of the type RR Lyrae), identified as a
thick disk subsystem by the kinematic criterion .


\section {BASIC DATA}

To analyze the behavior of certain chemical elements
in globular clusters, we took metallicity
from the computer version of the compilation catalog
(Harris~2010), spectroscopic determination of
the abundance of iron and the relative abundances of
two $\alpha$-elements --- titanium and calcium --- from our
compilation catalog (Marsakov~et\,al.~2019c). The
components of spatial velocities for stratification
by galactic subsystems of 115 clusters, defined by
Chemel~et\,al.~(2018) according to modern catalogs,
are given in Marsakov~et\,al.~(2019b). Analogous
data, as well as the relative abundances of one element of
slow neutron capture (yttrium), for 100 field variable
stars of the type RR Lyrae are taken from our catalog
\footnote{http://vizier.u-strasbg.fr/viz-bin/VizieR?
-source=J/AZh/95/54}
described in Marsakov~et\,al.~(2018). For comparison,
we used the catalog from Venn~et\,al. (2004), which
shows metallicity, relative abundances of $\alpha$-elements,
$s$-elements and components of spatial velocities for
785 stars of the galactic field in the entire range of
metallicity of interest to us, as well as the catalog in
Bensby et\,al.~(2014), which abundances similar data for
714 F--G~dwarfs and giants of the field, belonging
mainly to the disk subsystems of the Galaxy, that is,
to the sample of very few metal-poor stars. The errors
of the averaged relative abundances of two $\alpha$-elements
used in this paper are approximately the same for
clusters and lyrids of the field: 
$\langle\varepsilon\rm{[Ca, Ti/Fe]}\rangle \approx 0.11$,
and the errors of the spatial velocity components are
approximately 17~km\,s$^{-1}$ (for details, see Marsakov
et\,al. 2018, 2019b).

Fig.~1a,b shows the metallicity functions of globular
clusters and field variables of the type RR Lyrae
and, for comparison,--- two samples of nearby field
stars (1c,d). Within each distribution, the histograms
of the same objects that kinematically fall into the
thick disk subsystem are highlighted in darker color.
For more information on the separation of our field
lyrids and globular clusters to the subsystems of the
Galaxy, see Marsakov~et\,al.~(2018, 2019b). Despite
the fact that the method of stratification of all objects
is the same (that is, the components of the residual
velocities of all objects fall into the same ranges, the
values of which are set in advance), there is a difference
in the metallicity distributions between different
objects caught in a thick disk. The difference between
field stars and older objects --- globular clusters and
field stars of the type RR Lyrae is mainly noteworthy.
From Fig.~1 it is seen that all field stars (except
the lyrids) are on average more metallic. So, if the
field stars of the thick disk have average metallicities
$\rm{[Fe/H]} = -0.58 \pm 0.04$ and $-0.44 \pm 0.03$ according
to Venn~et\,al.~(2004) and Bensby~et\,al. (2014) respectively,
for clusters and field lyrids they are an order
of magnitude smaller --- $\langle\rm{[Fe/H]}\rangle = -1.16 \pm 0.11$ 
and $-1.39 \pm 0.04$ respectively.
Moreover, whereas in globular
clusters, the objects with the kinematics of the thick
disk fairly uniformly fill the entire range of metallicity,
the field lyrids show a confident maximum in the
region of $\rm{[Fe/H]} \sim -1.3$. In close field stars, this
maximum is observed at a much higher metallicity ---
$\rm{[Fe/H]} \approx -0.4
$\footnote{Note that the maxima of distributions in field stars of the
thick disk have slightly higher metallicity than the average
values due to long low-metal ``tails''.}. 
Let's consider the difference in the
chemical composition of different objects in more detail.

\section {GLOBULAR CLUSTERS} 

The results of Marsakov~et\,al.~(2019b,c) show
that the kinematic method of stratification is hardly
suitable for globular clusters of our Galaxy, since the
clusters of different subsystems selected by kinematics
radically differ in chemical properties from the field
stars of the same galactic subsystems. In particular,
all metal-rich ($\rm{[Fe/H]} > -1.0$) clusters that belong, according
to the kinematic criterion from Bensby et al.
(2003), to different subsystems, are within rather restricted
limits with respect to the center and plane
of the Galaxy. But in the range of less metallicity
among clusters with kinematics of the thick disk there
are also objects quite distant from the galactic plane,
not to mention clusters with kinematics of the halo.
At the same
time, among the metal-poor clusters that belong, by
kinematics, not only to the halo, but also to the thick
disk, there is a large number of them lying far beyond
the solar circle. This is reflected in the well-known
negative radial and vertical metallicity gradients in the
general population of globular clusters of the Galaxy.
As a result, it turns out that the traditionally used
procedure for separating globular clusters of the thick
disk from halo clusters by metallicity is more acceptable
(see the justification in Marsakov~et\,al.~2019b,c).

Let us consider in more detail the chemical composition
of globular clusters, kinematically related
to different subsystems, based on the study of the
relative abundances in globular clusters of only two $\alpha$-
elements --- calcium and titanium, as the most informative
for the diagnosis of the evolution of the early
Galaxy. In the visible range of the spectrum, these
two chemical elements have many lines, and their
abundances are fairly reliably determined. The choice
of these elements is due to the fact that the average
relative abundances of the two primary $\alpha$-elements ---
oxygen and magnesium --- in the course of the evolution
of a globular cluster are reduced in comparison
with their abundances in the parent proto-clouds.
And the abundances of another $\alpha$-element --- silicon --- are
defined for a smaller number of clusters and are not
defined at all for the field stars from Venn~et\,al.~(2004),
which we use for comparison. Although titanium,
strictly speaking, is not exclusively an $\alpha$-element and
partially belongs to the iron peak elements produced
in SNe Ia as well, its relative abundances well follow
the behavior of ``pure'' $\alpha$-elements -- O, Mg, Si, and
Ca (see, in particular, Bensby et al. 2014). The
same authors note that, unlike other $\alpha$-elements,
the uncertainty in [Ti/Fe] is low and unchanged for
almost all parameters. Therefore, they (and other
authors) most often use the results on titanium as
a more informative element to study the properties
of galactic subsystems. Fig.~2 shows the diagram
``metallicity -- relative abundances of $\alpha$-elements'' for field
stars and globular clusters. For both types of objects,
different icons indicate the belonging to different
galactic subsystems by kinematic criteria. The
darkest, most prominent icons indicate objects with
kinematics of the thick disk. It can be clearly seen
from the figure that the clusters of this subsystem
have not only a metallicity range different from the
range of field stars, but also relative abundances of $\alpha$-
elements significantly increased with the same metallicity
in the $\rm{[Fe/H]} > -1.0$. Moreover, in 7 out of 9
clusters, the relationship is $\rm{[Ca, Ti/Fe]} > 0.2$, and in
two (NGC 6528 and NGC 6553) only slightly less.
The three signed clusters at the bottom of the diagram
in the past belonged to destroyed dwarf galaxies
(Marsakov~et\,al.~2019c). In addition, only 9\,\% of field
stars of the thick disk are metal-poor ($\rm{[Fe/H]} < -1.0$),
whereas among the clusters there are half of such ones.  
Let us pay attention to one more feature of clusters. 
White small circles inside large circles in the 
figure indicate clusters
that we considered genetically related to a single
protogalactic cloud. These are clusters with a
direct revolution around the galactic center, which,
according to the positions and elements of the galactic
orbits, no one has associated with the destroyed
dwarf satellite galaxies, and all of them are located or
have maximum points of their orbits less than 15~kpc.
There are 60\,\% of such clusters in the sample. Among
them were clusters with the kinematics of all three
galactic subsystems identified by us, as well as all
metal-rich clusters (except for two with halo kinematics
and retrograde orbits).

It is not unlikely that the difference in the chemical
composition of globular clusters and field stars,
assigned to the subsystem of the thick disk by the
kinematic criterion, indicates the absence of a connection
between these subsystems of the same name.
Perhaps the reasons for the formation of similar subsystems
in field stars and globular clusters are different.
Indeed, as can be clearly seen from Fig.~2,
all metal-rich ($-1.0 < \rm{[Fe/H]} < 0.0$) clusters (except
for the accreted clusters Pal 12 and Ter 7) are located
in the upper part of the strip occupied by the
field stars. The calculation shows that the average
relative abundance of $\alpha$-elements and the error of
the mean of these clusters are $\langle\rm{[Ca, Ti/Fe]_{GC}}\rangle =
0.27 \pm 0.02$, whereas for field stars of the thick disk
in the same metallicity range beyond error is less:
$\langle\rm{[Ca, Ti/Fe]}_{field stars}\rangle = 0.14 \pm 0.01$. This difference
is clearly seen from the smoothed trends plotted for
globular clusters and field stars in Fig.~2. Recall
that according to modern concepts, $\alpha$-elements and
a small amount of iron are thrown into the interstellar 
medium by type II supernovae, whose lifetime
is less than 100~Myr. The main amount of iron
peak elements is produced during bursts of type Ia
supernovae, which begin to explode in mass approximately
1~Gyr after the start of star formation (Matteucci 2003).
Therefore, after the onset of mass bursts
of SNe Ia in the star-gas system, the so-called ``kink''
on the dependence ``[Fe/H] -- [$\alpha$/Fe]'' is observed. As
you can see from the smoothed trend for field stars
in Fig.~2 from Venn~et\,al.~(2004), in our Galaxy a
``kink'' is observed in the vicinity of $\rm{[Fe/H]} \approx -1.0$.
Hence, the high relative abundances of $\alpha$-elements in
metal-rich clusters indicate that they formed within
about a billion years after the start of star formation.
This means that metal-rich stars are younger than
globular clusters of the same metallicity.

The clusters identified by us as genetically related
are by definition located closer than 15~kpc from
the galactic center. In addition, as seen in Fig.~2 and 3,
they are the absolute majority in the metal-rich group,
which, on the metallicity distribution (see Fig. 1a), is
separated near the point $\rm{[Fe/H]} \approx -1.0$ by a dip from
less metallic clusters. Such properties of metal-rich
clusters can probably be explained by the existence
of an active phase in the Galaxy's evolution (see
Marsakov and Suchkov 1976, 1977, for more information).
The active phase period occurs after massive
supernova bursts in the halo, heating interstellar
matter, resulting in a delay in star formation. During
this delay, the interstellar matter of the Galaxy,
already contaminated with heavy elements, is mixed,
cools, and collapses to a smaller size, after which
star formation begins again and disk subsystems are
formed. However, this scenario of the formation of
subsystems in globular clusters contradicts, as can be
seen from Fig.~2, high relative abundances of $\alpha$-elements
in them: $\rm{[\alpha/Fe]} > 0.2$. As noted above, the high
[$\alpha$/Fe] ratios of almost all clusters indicate that they
all formed before the beginning of SNe Ia bursts, that
is, during the first billion years after the beginning of
star formation in the proto-galactic cloud. The same
figure shows that within the metallic range, clusters
can also show a decrease in the relative abundances of
$\alpha$-elements with an increase in metallicity, but on
average, the ratio [$\alpha$/Fe] for any metallicity in this
range remains higher than for field stars of the thick
disk. As a result, in the range $\rm{[Fe/H]} > -1.0$, the
dependence of [$\alpha$/Fe] on [Fe/H] for them is higher
and parallel to the same dependence for stars. And
among them there are clusters of all the subsystems
allocated by kinematics. A significant proportion of
such metal-rich clusters have the kinematics characteristic
of halo objects, that is, they are on very elongated
orbits (eccentricities up to $e = 0.9$), although
the orbits themselves are completely located inside
the solar circle (see Fig.~2 and 3). This could well have
happened during the formation of these clusters during
the renewed collapse of the proto-galactic cloud,
already enriched with heavy elements, after the active
phase.

If the appearance of metal-rich clusters that are genetically
linked to a single proto-galactic cloud owes
its origin to the active phase, then all of them must
be younger than less metallic clusters. Indeed, since
metallicity increases over time in our Galaxy from the
halo to the disk, it is therefore a statistical indicator
of age. This is clearly seen in Fig. 3, where the
relationship of metallicity to age is given for globular
clusters from VandenBerg~et\,al.~(2013). In the figure,
genetically related clusters are highlighted in white
circles. We see that all metal-rich clusters are younger
than 11.5 Gyr (47 Tuc is only slightly older, and the
old cluster NGC 6362, as seen in Fig.~1a, is slightly
to the left of the dip in the histogram, i.e. in the metal-poor
range), while all less metallic, genetically related
clusters are older (except NGC 362). However,
according to the definitions from Salaris and Weiss
(2002), all metal-rich clusters are older than this age and
appeared simultaneously with the oldest 
clusters of least metallicities, 
which indicates that the age of clusters is not
sufficiently reliable. From Fig.~3 one can see a steady
monotonous decrease in the ages of a complete sample
of clusters with an increase in metallicity. In this
case, two separate parallel dependences for clusters
of varying metallicities are clearly distinguished, with
age ranges for these dependences being the same,
while their [Fe/H] ranges are, respectively: from $-2.4$
to $-1.2$~dex and from $-1.3$ to $-0.3$~dex. We see that
the more metallic sequence consists entirely of genetically
related clusters (except NGC 6652). About
half of the genetically related clusters turned out to
be in a less metallic group. Note that the authors of
VandenBerg et\,al.~(2013) themselves did not find an
unambiguous explanation of the nature of the two sequences,
but attributed their occurrence to the difference
in the clusters' loss of gas ejected by their giants
of the asymptotic branch --- it is from this gas that the
second generation of cluster stars with the changed
chemical composition is born afterwards. From Fig. 3
it is also seen that clusters of only one sequence fall
into the range $\rm{[Fe/H]} > -1.0$. Recall that we included
all close clusters ($R_{G} < 15$ kpc) in the genetically
related ones, not all of which have computed orbital
elements, so there may be distant ones (i.\,e., with
$R_{max} > 15$~kpc), but formally such clusters cannot be
assigned to this group. Note that the dip we have
discussed in the metallicity function in the vicinity
$\rm{[Fe/H]} \approx -1.0$ can hardly be considered a consequence
or cause of the existence of two sequences of
clusters with different ages and metallicities in Fig.~3,
because, although it cuts off the metal-poor sequence,
it divides the metal-rich one almost in half. Note that in
Fig.~3 there is only one equally young, but metal-poor,
genetically connected cluster, whose orbit also lies
completely inside the solar circle --- NGC 362.

It turns out that the most metal-rich, genetically
related clusters were born later. In the lower right
quadrant of Fig.~3 in the metallic range you can see a
statistically insignificant, weak tendency to decrease
age with increasing metallicity. At least two of the
youngest metal-rich clusters belong to a thin disk
by kinematics. By determination from Salaris and
Weiss (2002), this dependence is also very uncertain.
Apparently, for a final clarification of the question
of the behavior of the ages of clusters in the range
$\rm{[Fe/H]} > -1.0$, their new determinations are needed
with modern photometric and astrometric data. Only
two (NGC 6652 and NGC 6637) of the metal-rich
clusters in Fig.~3 have the halo kinematics. Moreover,
the second cluster has an orbit completely inside the
solar circle. According to the ages we used from
VandenBerg~et\,al.~(2013), the supernovae SNe Ia still
manage to enrich the interstellar medium with metals,
from which subsequently clusters of the metal-rich group
were born. In this case, the chemical and kinematic
properties of metal-rich clusters do not contradict the
hypothesis of active phases in the evolution of the
Galaxy. Moreover, the higher relative abundances of 
$\alpha$-elements in all metal-rich clusters can be explained
by the fact that they were formed from high-density
interstellar matter enriched with heavy elements after
a delay in star formation, which led to an increase
in the upper limit of the masses of the formed stars,
and consequently, of supernovae of the second type,
ejecting a greater number of $\alpha$-elements. Then metal-poor,
genetically related clusters, whose orbits also
lie almost entirely within the solar circle, should have
been born before the onset of the active phase. Indeed,
as seen in Fig.~3, the ages of less metallic genetically
related clusters are systematically greater. It is possible
that by the time they were born, the rate of collapse
of the proto-galactic cloud slowed significantly, which
led to the appearance of clusters among them with a
``younger'' kinematics of the thick disk.

Thus, using the hypothesis of active phases in the
evolution of the Galaxy, we can try to give a consistent
explanation of the reason for the abrupt change in
the volume of the Galaxy occupied by clusters when
passing through $\rm{[Fe/H]}\approx -1.0$. Nevertheless, the
existence of old metal-poor clusters with the kinematics
of the thick disk and with orbits also enclosed
within the solar circle leaves open the question of the
reason for the divergence of stratification results on
the metallicity and kinematics of clusters and field
stars. It turns out that what we call the thick disk
of field stars and clusters are different subsystems.

Otherwise, if subsequently it turns out that there is
no dependence between metallicity and age for globular
clusters, a consistent explanation cannot be found
why, when passing through $\rm{[Fe/H]} \approx -1.0$, clusters
abruptly change the occupied volume in the Galaxy.
But in any case, the high relative abundances of 
$\alpha$-elements in metal-rich clusters indicate their greater
age than the age of field stars of the thick-disk, in
which these abundances are much smaller due to the
massive pollution of their parent interstellar medium
by emissions of relatively long-lived SNe Ia.

\section {FIELD VARIABLE STARS of TYPE RR Lyrae}

The contradiction, although not so pronounced,
between the criteria of belonging to disk subsystems
and to halo in terms of chemical and kinematic
properties is also observed between two types of field
stars: F--G dwarfs and giants on the one hand, and
variables of the type RR Lyrae, on the other (for more
information, see Marsakov et\,al.~2018). Such variable
stars are typical representatives of globular clusters,
so the similarity of their metallicity functions is not
surprising. From Fig.~1 it can be seen that the
percentage of metal-poor objects with thick disk kinematics
in the field lyrids is even higher than in globular
clusters. But, unlike clusters, the relative abundances of
$\alpha$-elements in metal-rich lyrids systematically decrease
with the growth of [Fe/H], as in other stars of the
field. This is clearly seen in Fig.~4, where the diagram
``metallicity -- relative abundances of $\alpha$-elements'' for field
stars and field variables of type RR Lyrae is shown.
The abundances of the same two $\alpha$-elements --- calcium
and titanium --- are taken from the author’s compilation
catalog described in Marsakov~et\,al.~(2018). In
this paper, we also proposed a possible explanation
for these differences by the fact that, being older
stars than the main part of dwarfs, lyrids track the
chemical composition of the interstellar medium at
the initial stages of the formation of the thick disk
subsystem. Unfortunately, the ages of these variables
cannot be determined, as in the case of globular
clusters. A knee in the dependences of [$\alpha$/Fe] and
[Fe/H] in the lyrids indicates the fact that the era of
type Ia supernovae outbursts has begun in the star-gas system
in which they were formed, that is, about 1~Gyr have
passed since the start of star formation. The large
duration of the evolution of the thick disk subsystem
is also clearly observed in Fig.~1c,d and Fig.~3c,d from
Marsakov~et\,al.~(2018) systematic trends within this
subsystem of both metallicity and relative abundances of
$\alpha$-elements with changes in kinematic indicators.

Pay attention to the peculiarities of the chemical
composition of the lyrids observed in Fig.~4, which
can be associated with the difference in the nature of
some field stars. In particular, the recent work Mackereth
et\,al. (2019) analyzed the relative abundances of 
$\alpha$-elements and velocities of several tens of thousands
of stars within 15 kpc from the Sun. The sample was
compiled by cross-identification between the SDSS--APOGEE 
DR 14 and Gaia DR2 catalogs. As a result,
it was concluded that, in the early stages of evolution,
our Galaxy captured a massive (about $10^{9}M_{\odot}$) satellite
galaxy, as a result of which part of the field stars
born in this satellite galaxy fell into our Galaxy, and
part of the stars of the already formed thin disk thus
``heated'', forming a subsystem of the thick disk.
We emphasize that Mackereth et\,al.~(2019) notes that
the nature of such low-velocity stars is not completely
clear, and the authors only assume that the thick disk
subsystem was formed as a result of simultaneous
processes --- star formation in a single protogalactic
cloud and accretion processes. This assumption is
also supported by the conclusions of Belokurov et al.
(2018), the authors of which investigated the change
in the relationship between the Oosterhof classes of
field stars of the type RR Lyr with distance from the
galactic center according to GAIA data. They found
confirmation of the fact that some of these stars entered
our Galaxy from a decaying dwarf satellite galaxy
of large mass. The same conclusion was reached by
Helmi~et\,al.~(2018), who, basing on APOGEE and
Gaia DR2, as well as numerical simulations, showed
that remnants of a dwarf galaxy (they called it Gaia-Enceladus) 
more massive than the Small Magellanic
Cloud prevail in the inner halo. They demonstrated
that among the stellar objects they studied, hundreds
of lyrids and more than a dozen globular clusters
formed in this galaxy. Moreover, in their opinion,
the merging with Gaia-Enceladus, led to a dynamic
``heating'' of the predecessor of the thick disk
of the Galaxy and, consequently, contributed to the
formation of this component about 10 Gyr ago.

This means that not all of the field lyrids are
genetically related to our Galaxy. Indeed, by now,
some of these ``captured'' field stars may well have
become variables of the type RR Lyrae. Unfortunately,
due to the lack of observational data, it is
not possible to trace the existence of a ``knee'' in
our field lyrids with a known abundance of chemical
elements. However, from the form of smoothed trends
on the dependence of [Ca,Ti/Fe] on [Fe/H] (Fig. 4a)
it can be seen that starting from $[Fe/H] \approx -1.3$ the
sequence of lyrids is lower than that of the field stars,
and this value coincides with the metallicity of the
``knee'' noted in Mackereth~et\,al.~(2019) for accreted
Galaxy stars according to APOGEE data. This
fact, as well as the relatively lower relative abundances
of $\alpha$-elements, especially titanium (see Fig.~2 in
Marsakov~et\,al.~2018), in most metal-rich lyrids,
with the kinematics of thick and thin disks in Fig.~4
compared with nearby stars, may testify in favor of
the extragalactic origin of some, in particular metal-rich
lyrids. As the calculation showed, in the range
($\rm{[Fe/H]} > -0.5$), all lyrids on average have ratios
$\langle\rm{[Ca, Ti/Fe]}_{RRLyr}\rangle = -0.02 \pm 0.03$, 
which is far beyond errors and lower than in field 
stars of equally metallicity: 
$\langle\rm{[Ca, Ti/Fe]}_{field stars}\rangle = + 0.042 \pm 0.003$.
In Fig.~4a it can be seen that almost all these lyrids
lie in the lower part of the strip occupied by the field
stars$\footnote{Features of the change in each of the 
four $\alpha$-elements can be seen in Fig.~2a-d in 
Marsakov~et\,al.~(2018), which show that the lowest 
relative contents are observed for titanium,although 
for other elements they are usually lower than the
average for field stars.}$.

It is known that, along with $\alpha$-elements, supernovae
of the type SNe II also eject atoms of elements
of $r$-processes, in particular, europium. Therefore,
the relative abundances of this element, which is well defined
in stars, change depending on the metallicity like
$\alpha$-elements, which allows increasing the reliability of
determining the course of the dependence ``[el/Fe] --
[Fe/H]'' from a single chemical element. But, unfortunately,
neither europium nor other $r$-elements were
found for any field lyrids with $rm{[Fe/H]} > -1.0$. Although
for other metal-rich variable stars --- Cepheids ---
the abundances of europium are well defined and the
relations [Eu/Fe] are the same as for field dwarfs and
giants (Marsakov~et\,al.~2013).

However, not only the [Ca,Ti/Fe] ratios, but also
the relative abundances of the light $s$-element --- yttrium
--- demonstrate very low values in metal-rich lyrids. This
is clearly visible in Fig.~4b, where the change in the
[Y/Fe] ratio with the change in metallicity for the
field lyrids and comparison stars is shown. Note
that heavy $s$-elements in metal-rich lyrids give almost
solar [el/Fe] ratios. Moreover, according to modern
concepts, the overwhelming number of atoms of all
$s$-elements is produced in the interior of the giants
of the asymptotic branch with masses greater than
$4 M_{\odot}$ (the main component of the $s$-process) and, by
dumping the shell, enters interstellar space. Recall
that some stars of the asymptotic branch of the giants
turn out to be close binaries that explode later as
type Ia supernovae. That is, the time of ejection of
these elements and iron are approximately the same.
Nevertheless, in the metallic range in Fig. 4b, together
with a sharp jump-like decrease in the [Y/Fe]
ratios for lyrids, one can see a slight tendency to increase
them with increasing [Fe/H], as for stationary
field stars. It turns out that metal-rich lyrids are formed
from a substance depleted in this element. We emphasize
that the low [Y/Fe] ratios in the field lyrids
were previously noticed and people tried to explain
them by the unusual state of the atmospheres of these
variable stars. In particular, Clementini~et\,al.~(1995)
suggests that the abnormal abundance of yttrium is
caused by superionization caused by strong emission
lines Ly$\alpha$, which are induced by shock waves in a
pulsating atmosphere. However, Liu~et\,al. (2013)
denied this possibility due to the lack of a similar
effect in metal-poor lyrids and proposed to consider
this an effect caused by the difference in surface gravity
between evolved stars of the type RR Lyrae and
non-evolved dwarfs. However, numerous definitions
of gravity accelerations on the surface of metallic
lyrids from high-resolution spectra showed that they
are in the range: lg $g = (2.5 - 3.0)$, that is, they are
approximately the same as in our comparison giants
(see, for example, Marsakov~et\,al.~2019a). And in
other metal-rich stars --- Cepheids --- gravitational accelerations
are even less than that of lyrids, and the
[Y/Fe] ratios are slightly larger than that of dwarfs
and giants of the field (see Andrievsky~et\,al.2013). Since
the influence of atmospheric features cannot explain
the abnormal yttrium abundance in metal-rich lyrids, it can
be assumed that it is caused by an external cause.
Perhaps in the interstellar medium, from which metal-rich stars
were formed, which have now become lyrids, 
in addition to an excess of helium, there were also a deficit 
of $\alpha$-elements and the light $s$-element ---
yttrium? The fact is that in Marsakov~et\,al.~(2019a)
we hypothesized that relatively young metal-rich field
lyrids have increased helium abundances leading to faster
evolution of stars, and in the vicinity of the Sun they
are carried out by radial migration from the central
regions of the Galaxy, where such stars are already
detected. If we now assume that part of the field lyrids 
has an extragalactic origin, it becomes necessary to
explain how such metal-rich stars could have been
formed in the early stages of evolution, now in a dwarf
galaxy, and in our Galaxy to acquire the kinematics of
field stars of the thick and thin disk. Of course, this
assumption is very superficial and requires a comprehensive
justification.

\section {CONCLUSIONS} 

Thus, the differences in the abundances of certain
chemical elements in two representatives of the old
stellar populations of the Galaxy --- globular clusters
and field variables of the type RR Lyrae, -- from similar
abundances in field dwarfs and giants, which by kinematic
parameters belong to the thick galactic disk,
may indicate their formation from matter that has
passed through different paths of chemical evolution.
As a result, we can assume that the thick disk subsystem
in the Galaxy turns out to be composite, and
at least three components exist independently within
it. The oldest one is metal-rich globular clusters, which
were formed from a single protogalactic cloud shortly
after the onset of bursts of type Ia supernovae in it.
Then a subsystem of field stars of the thick disk was
formed as a result of the ``heating'' of the stars of the
thin disk already formed in the Galaxy by a rather
massive dwarf satellite galaxy that fell on it. And,
finally, a subsystem of field stars with the kinematics
of not only a thick, but even a thin disk that fell on
the Galaxy from this captured satellite galaxy. In this
work, by the representatives of the last subsystem,
we consider metal-rich field variable stars of the type
RR Lyrae. At present, we are preparing an article for
publication in which we will demonstrate on basis of
more voluminous observational material, the differences
in abundances of chemical elements, other than
those considered, in variable stars of type RR Lyrae
and stationary field stars, and also will analyze the
possible causes of their occurrence.

\section*{ACKNOWLEDGMENTS}

The authors are grateful to the anonymous referee
for valuable comments that made the results of the
work more reasonably presented.

\section*{FUNDING}

Authors are grateful to the Laboratory of Cosmic 
Microphysical Studies of the Structure and Dynamics 
of the Galaxy, Institute of Physics, Southern 
Federal University, for the support.

\section*{CONFLICT OF INTEREST}

The authors declare no conflict of interest.

\renewcommand{\refname}{REFERENCES}

\newpage

\begin{figure*}
\centering
\includegraphics[angle=0,width=0.99\textwidth,clip]{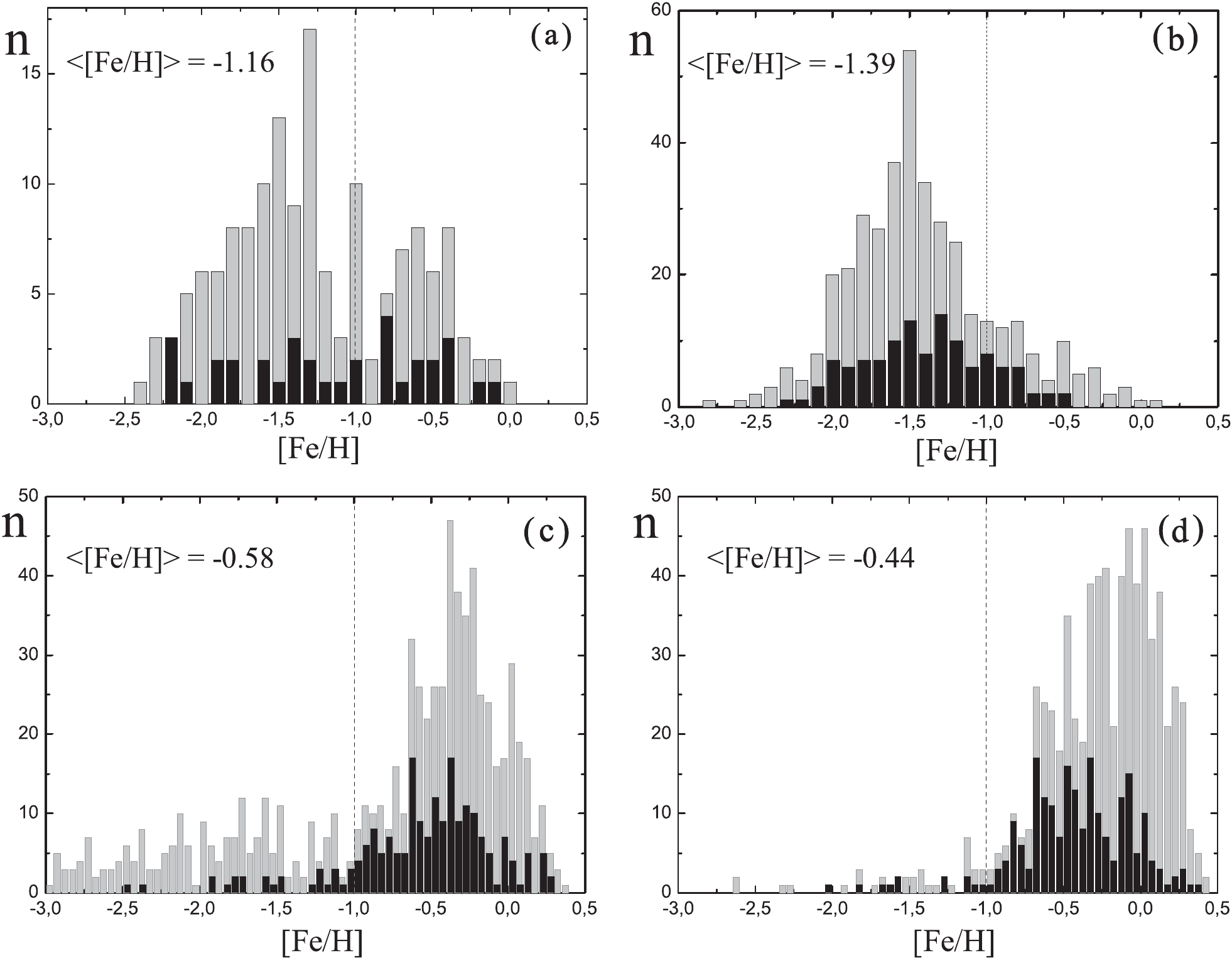}
\caption{Metallicity distributions of globular clusters (a), 
         field lyrides (b), field stars from Venn~et\,al.~(2004) (c), 
         and F\,--G --field stars from Bensby~et\,al.~(2014) (d). 
         Distributions of objects with the thick disk kinematics 
         are highlighted in dark color. The metallicities for 
         clusters are taken from Harris (2010), for lyrids 
         from Dambis~et\,al.~(2013), and the spectroscopic 
         values [Fe/H] for field stars -- from the above articles. 
         Vertical dashed lines on all panels at $\rm{[Fe/H]} = -1.0$ 
         roughly correspond to a dip or bend in the histograms 
         for all objects. The panels show the average 
         metallicity of objects of the thick disk.}
\label{fig1}
\end{figure*}

\newpage

\begin{figure*}
\centering
\includegraphics[angle=0,width=0.99\textwidth,clip]{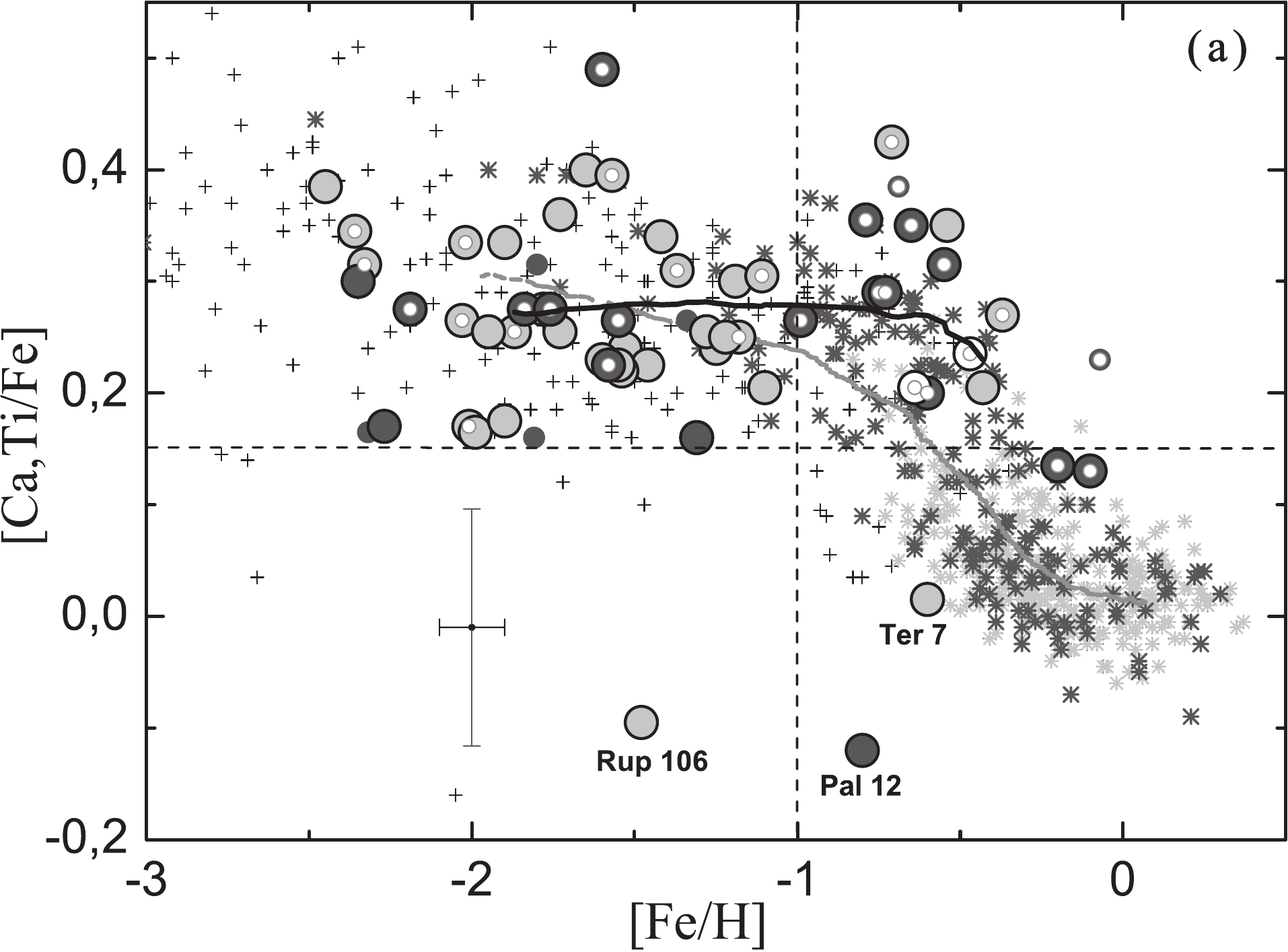}
\caption{Changes in the relative contents averaged over two 
         $\alpha$-elements (Ca and Ti) with a change 
         in the metallicity of field stars from Venn~et\,al.~(2004) 
         and globular clusters from Marsakov~et\,al.~(2019b). 
         The field stars are indicated as follows: light gray 
         snowflakes are for the thin disk, dark snowflakes 
         are for the thick disk, dark crosses are for halo. 
         Large circles of the clusters, which, according 
         to kinematic features, belong to a thin disk are 
         light, to a thick disk--dark, halo--gray. Small 
         dark circles denote non-stratified clusters. 
         White circles inside large circles denote genetically 
         related clusters. Three clusters lost by dwarf 
         satellite galaxies are signed. The dark and gray 
         broken curves are smoothed trends with a moving 
         average for globular clusters and field stars, 
         respectively. The average error bars of individual 
         definitions are plotted. The vertical dashed line 
         $\rm{[Fe/H]} = -1.0$ divides the clusters into 
         two groups metal-poor and metal-rich, and the 
         horizontal line [Ca,Ti/Fe]= 0.15 is approximately 
         drawn along the lower boundary between metal-poor 
         globular clusters and field stars.}
\label{fig2}
\end{figure*}

\newpage

\begin{figure*}
\centering
\includegraphics[angle=0,width=0.99\textwidth,clip]{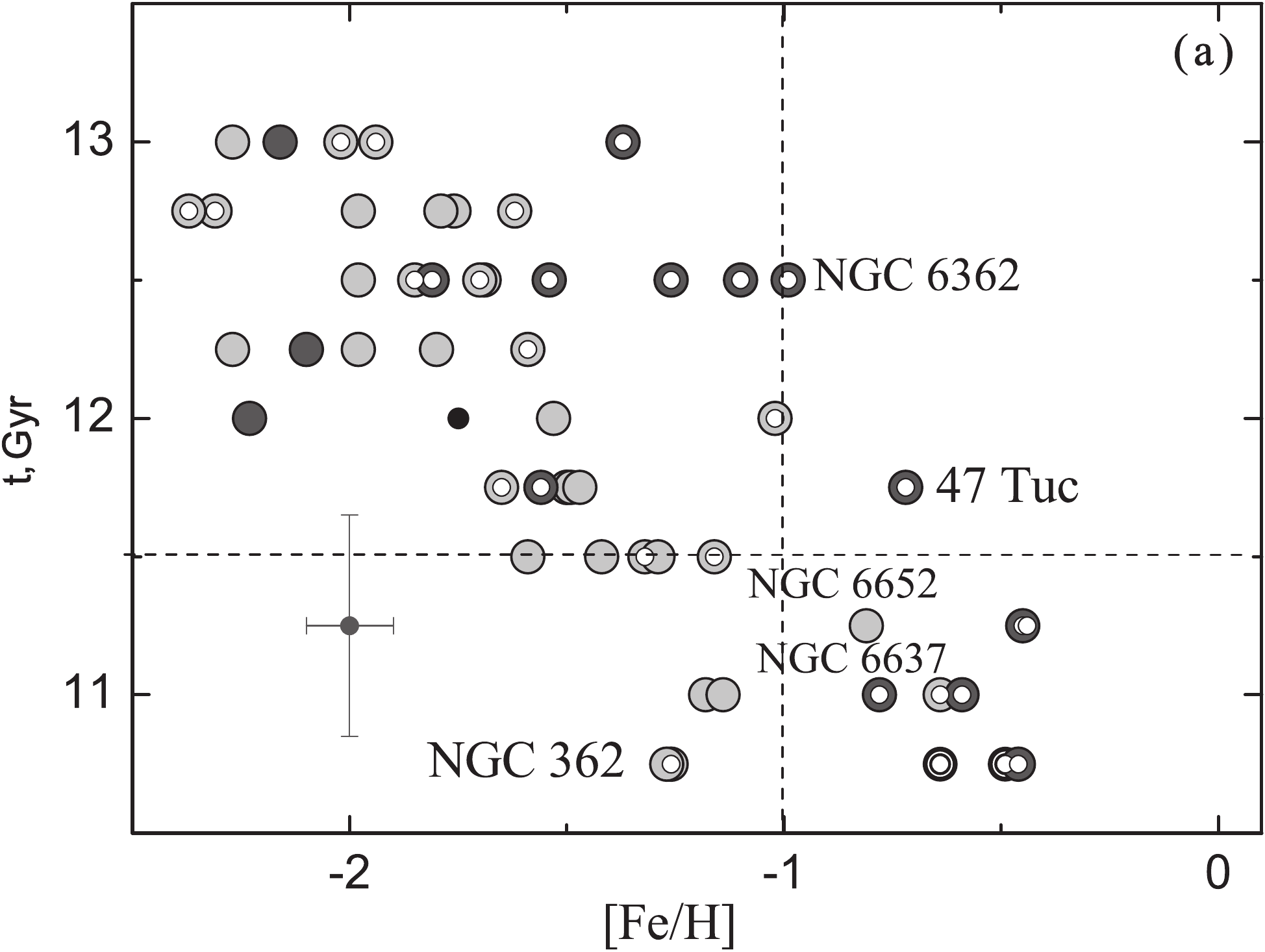}
\caption{Metallic -- diagram for all globular clusters. 
         Metallicities are taken from Harris (2010), and ages 
         from VandenBerg et\,al.~2013). Designations, as 
         in Fig.~2 The clusters that are mentioned in the 
         text are signed. The vertical dashed line 
         approximately corresponds to a dip in the 
         metallicity function, and the horizontal 
         dashed line corresponds to an age value of 11.5~Gyr.}
\label{fig3}
\end{figure*}

\newpage

\begin{figure*}
\centering
\includegraphics[angle=0,width=0.99\textwidth,clip]{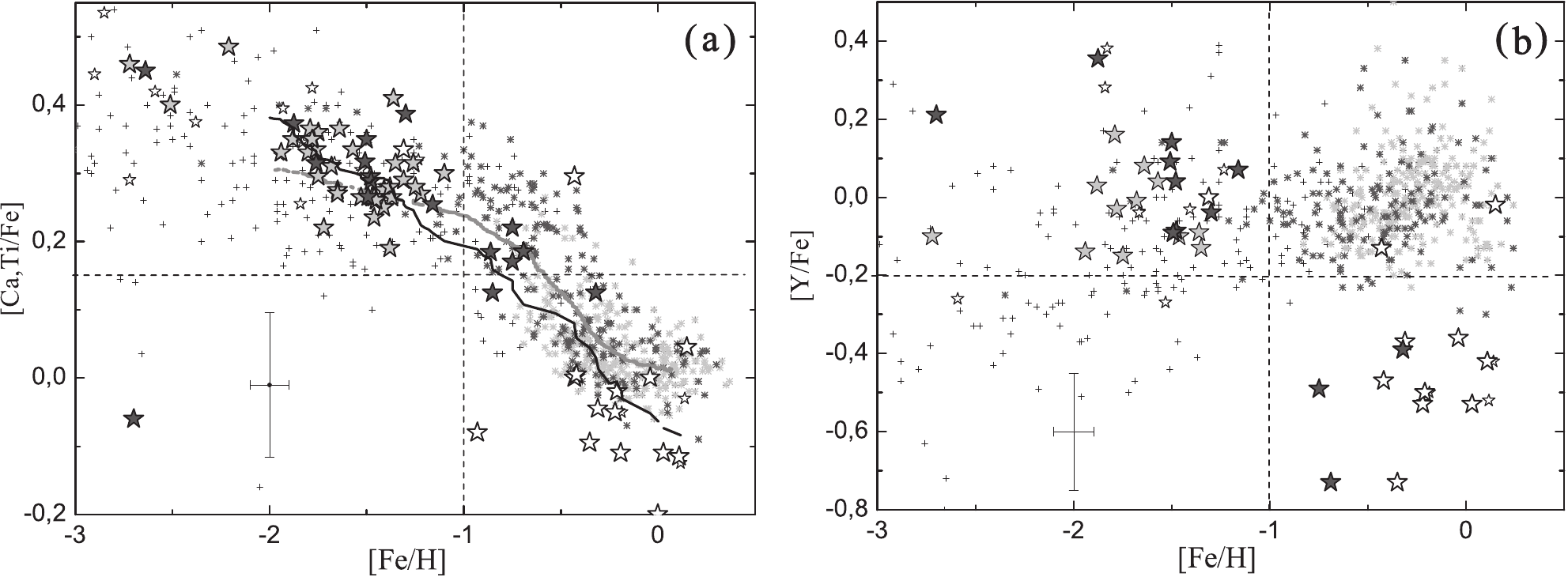}
\caption{Changes in the relative abundances averaged over two 
         $\alpha$-elements (Ca and Ti) (a) and the relative  
         of yttrium (b) with a change in metallicity for 
         field stars from Venn~et\,al.~(2004) and field 
         stars of the RR Lyrae type from Marsakov~et\,al.~(2018). 
         Notations are identical to those of Fig.~2, but the 
         lyrids are marked with asterisks of the corresponding 
         color. The dark and gray broken curves are the 
         smoothed trends for lyrids and field stars, 
         respectively (a). On both panels, the vertical 
         dashed lines are as in Fig.~2, and the horizontal 
         ones approximately correspond to the lower values 
         of the corresponding [el/Fe] ratios for metal-poor lyrids.}
\label{fig3}
\end{figure*}



\begin{thebibliography}{}


\bibitem[1.]{1.} M. G. Abadi, J. F. Navarro, M. Steinmetz, and V. R. 
Eke, Astrophys. J. 597 (1), 21 (2003). 

\bibitem[2.]{2.} S.M. Andrievsky, J. R. . D. L\'epine, 
S. A. Korotin, et al., Monthly Notices Royal Astron. Soc. 428 (4), 3252
(2013).

\bibitem[3.]{3.} V. Belokurov, D. Erkal, N. W. Evans, et al., Monthly
Notices Royal Astron. Soc. 478 (1), 611 (2018).

\bibitem[4.]{4.} T. Bensby, S. Feltzing, and I. Lundstr oЁ m, Astron. and
Astrophys. 410, 527 (2003). 

\bibitem[5.]{5.} T. Bensby, S. Feltzing, and M. S. Oey, Astron. and
Astrophys. 562, A71 (2014).

\bibitem[6.]{6.} T. V. Borkova and V. A. Marsakov, Astronomy Reports
44 (10), 665 (2000).

\bibitem[7.]{7.} C. B. Brook, D. Kawata, B. K. Gibson, and K. C.
Freeman, Astrophys. J. 612 (2), 894 (2004).

\bibitem[8.]{8.} A. A. Chemel, E. V. Glushkova, A. K. Dambis, et al.,
Astrophysical Bulletin 73 (2), 162 (2018).

\bibitem[9.]{9.} G. Clementini, E. Carretta, R. Gratton, et al., 
Astron. J. 110, 2319 (1995).

\bibitem[10.]{10.} A. K. Dambis, L. N. Berdnikov, A. Y. Kniazev, et al.,
Monthly Notices Royal Astron. Soc. 435 (4), 3206 (2013).

\bibitem[11.]{11.} O. J. Eggen, D. Lynden-Bell, and A. R. Sandage,
Astrophys. J. 136, 748 (1962).

\bibitem[12.]{12.} R. G. Gratton, E. Carretta, F. Matteucci, and C. Sneden,
Astron. and Astrophys. 358, 671 (2000).

\bibitem[13.]{13.} W. E. Harris, arXiv e-prints arXiv:1012.3224 (2010).

\bibitem[14.]{14.} A. Helmi, C. Babusiaux, H. H. Koppelman, et al.,
Nature 563 (7729), 85 (2018).

\bibitem[15.]{15.} T. D. Kinman, Monthly Notices Royal Astron. Soc.
119, 538 (1959).

\bibitem[16.]{16.} P. Kroupa, Monthly Notices Royal Astron. Soc.
330 (3), 707 (2002).

\bibitem[17.]{17.} B. V. Kukarkin, Gobular star clusters. The general
catalogue of globular star clusters of our galaxy,
concerning information on 129 objects known before
1974 (1974). In Russian.

\bibitem[18.]{18.} S. Liu, G. Zhao, Y.-Q. Chen, et al., Research in Astronomy
and Astrophysics 13 (11), 1307-1329 (2013).

\bibitem[19.]{19.} J. T. Mackereth, R. P. Schiavon, J. Pfeffer, et al.,
Monthly Notices Royal Astron. Soc. 482 (3), 3426 (2019).

\bibitem[20.]{20.} V. A. Marsakov, M. L. Gozha, and V.\,V.~Koval', Astronomy
Reports 62 (1), 50 (2018).

\bibitem[21.]{21.} V. A. Marsakov, M. L. Gozha, and V.\,V.~Koval', Astronomy
Reports 63 (3), 203 (2019a).

\bibitem[22.]{22.} V. A.Marsakov, V.\,V.~Koval', and M. L. Gozha, Astrophysical
Bulletin 74 (4), 403 (2019b).

\bibitem[23.]{23.} V. A. Marsakov, V.\,V.~Koval', and M. L. Gozha, Astronomy
Reports 63 (4), 274 (2019c).

\bibitem[24.]{24.} V. A. Marsakov, V.\,V.~Koval', V. V. Kovtyukh, and 
T.V. Mishenina, Astronomy Letters 39 (12), 851 (2013).

\bibitem[25.]{25.} V. A. Marsakov and A. A. Suchkov, Soviet Astronomy
Letters 2, 148 (1976).

\bibitem[26.]{26.} V. A. Marsakov and A. A. Suchkov, Sov. Astron. 21,
700 (1977).

\bibitem[27.]{27.} F. Matteucci, Astrophys. and Space Sci. 284 (2), 539
(2003).

\bibitem[28.]{28.} W. W. Morgan, Astron. J. 64, 432 (1959).

\bibitem[29.]{29.} J. X. Prochaska, S. O. Naumov, B. W. Carney, et al.,
Astron. J. 120 (5), 2513 (2000).

\bibitem[30.]{30.} M. Salaris and A. Weiss, Astron. and Astrophys. 388,
492 (2002).

\bibitem[31.]{31.} R. Sch\"onrich and J. Binney, Monthly Notices Royal
Astron. Soc. 396 (1), 203 (2009).

\bibitem[32.]{32.} D. A. VandenBerg, K. Brogaard, R. Leaman, and
L. Casagrande, Astrophys. J. 775 (2), 134 (2013).

\bibitem[33.]{33.} K. A. Venn, M. Irwin, M. D. Shetrone, et al., 
Astron. J. 128 (3), 1177 (2004).

\bibitem[34.]{34.} R. Zinn, Astrophys. J. 293, 424 (1985).




\end{thebibliography}
\end{document}